\documentclass[pre,showpacs,preprint]{revtex4}

\usepackage{graphicx}
\usepackage{natbib}
\usepackage{amssymb}
\usepackage{amsmath}
\usepackage{verbatim}

\begin{document}

\title{Dynamical heterogeneity in aging colloidal glasses of Laponite}

\author{S. Jabbari-Farouji $^{1,2}$, R. Zargar $^{2}$, G H.\ Wegdam $^{2}$ and Daniel Bonn$^{2,3}$}

\affiliation{$^{1}$ LPTMS, CNRS and Universit$\acute{e}$ Paris-Sud, UMR8626, Bat. 100, 91405 Orsay, France}
\affiliation{$^{2}$ Van der Waals-Zeeman Institute, Institute of Physics (IoP) of the Faculty of Science (FNWI) University of
Amsterdam,  1098 XH Amsterdam, the Netherlands}
\affiliation{$^{3}$ Laboratoire de Physique Statistique de l'ENS, 24 rue Lhomond, 75231 Paris Cedex 05, France}
\date{\today}

\date{\today}

\begin{abstract}
Glasses behave as solids due to their long relaxation time; however the origin of this slow response remains a puzzle. Growing dynamic length scales due to cooperative motion of particles are believed to be central to the understanding of both the slow dynamics and the emergence of rigidity.  Here, we provide experimental evidence of a growing dynamical heterogeneity length scale that increases with increasing waiting time in an aging colloidal glass of Laponite. The signature of heterogeneity in the dynamics follows from dynamic light scattering measurements in which we study both the rotational and translational diffusion of the disk-shaped particles of Laponite in suspension. These measurements are accompanied by simultaneous microrheology and macroscopic rheology experiments. We find that rotational diffusion of particles slows down at a faster rate than their translational motion. Such decoupling of translational and orientational degrees of freedom finds its origin in the dynamic heterogeneity since rotation and translation probe different length scales in the sample. The macroscopic rheology experiments show that the low frequency shear viscosity increases at a much faster rate than both rotational and translational diffusive relaxation times.
\end{abstract}

\maketitle

\newpage

\section{Introduction}
Glasses are non-equilibrium solids with no long-range structural correlations. Due to their long relaxation time, glasses behave as solids on experimental time scales. This slow response is often attributed to cooperative motion of particles in which a subregion of the liquid relaxes to a new local configuration \cite{adam_gibbs65,cooperative}. However, the size of cooperatively rearranging regions has never been observed to exceed a few particle diameters, and the observation of long-range correlations that are signatures of an elastic solid has remained elusive.  The universal dynamical feature in a variety of such systems, including molecular, colloidal, metallic is a dramatic slowing down of the diffusion of particles concomitant with huge increase of viscosity upon approaching the glass-transition \cite{Cates,Biroli-glasses}. The slowing down of dynamics expresses itself, among other things, in the non-exponential relaxation of translational and rotational degrees of freedom and non-decaying plateaus of the associated correlation functions in the glassy phase.

Due to the intrinsic non-equilibrium nature of all glassy systems, they exhibit aging. Aging means that the physical properties of the system evolve with waiting time i.e., the time elapsed since the quench into the glassy phase. The aging is often attributed to local rearrangements, i.e., locally an energy barrier can be overcome and the system falls into a new configuration in phase space. Cooperative motion requires spatially and or temporarily correlated local rearrangements, therefore suggests the existence of a  heterogenous  dynamics for the particle motion in both space and time.

Indeed, various theoretical and experimental studies have provided evidence for the existence of dynamical heterogeneity in both molecular and colloidal glass formers \cite{decoupDynHet, chandler, DynH} despite the differences  between the two systems. The building blocks of molecular glass formers are in the range 0.1-1 nm and the glass transition is these systems is generally driven by a temperature or pressure quench. Colloidal glasses, on the other hand,  consist of large particles with sizes in the range 10-1000 nm that undergo Brownian motion in a suspending liquid. The key control parameter  in colloidal systems is the volume fraction or density. These systems have  the advantage  that they can  be viewed directly using microscopy, so  that we can have a microscopic picture of how particles rearrange at the glass transition \cite{Weeks}.
In computer simulations \cite{chandler, Berthier} and confocal microscopy experiments on colloidal glasses \cite{Weeks,OrientationalG, Antina,Vijey,weeks1,Abou} the heterogenous regions were visualized directly. One way of quantifying  the spatio-temporal correlations is to determine  the four-point dynamic correlation functions  that measure the correlations of motion in space-time  \cite{Franz}. Such four-point correlation functions have been used to quantify the dynamical heterogeneity and its associated length-scale in several glass formers \cite{growingL,weeks1}; however this correlation function is not always easily accessible in measurements of the glass-forming systems, especially for molecular glass-formers and nanometer sized colloidal particles. Therefore, other probes that reveal the nature of  heterogenous dynamics are helpful.

One important consequence of dynamical heterogeneity is the break-down of mean-field dynamical relations such as the well-known Stokes-Einstein relation \cite{SE1,SE2,SE3,SE4, Biroli-SE}. If  spatially correlated mobile regions exist, it seems plausible that there should be a difference between the globally measured viscosity that results from the cooperative motion and the local diffusion dynamics of the particles that probe the local 'viscosity'. Therefore, a scale-dependent diffusion coefficient can be considered as a signature of dynamical heterogeneity and provides us with some indication of the dynamic correlation length scale.

 The translational diffusion coefficient and rotational relaxation time  of particles suspended in a (low-viscosity) liquid follow the Stokes-Einstein and Debye-Stokes-Einstein equations, respectively, that relate the translational diffusion coefficient and rotational relaxation time to the temperature and  the viscosity  as follows \cite{SE,DSE}:
\begin{eqnarray}\label{eq:eq1}
 D_{tr}&=&\alpha  \frac{k_BT}{ \eta R} \quad\quad\quad  \mathrm{Stokes-Einstein \quad relation} \\
\tau_r&=& \alpha'  \frac {\eta R^3} {k_BT} \quad\quad\quad\mathrm{ Debye-Stokes-Einstein \quad relation}
\end{eqnarray}
Here $R$ is the size of the largest dimension of the particles and $\alpha$ and  $\alpha'$ are prefactors that depend on the shape of particles and are known in principle. These relationships predict that translational and orientational relaxation times should be directly proportional to the macroscopically measured viscosity.

Here, by studying the aging dynamics of  relaxation time and viscosity  of a colloidal glass of Laponite \cite{glass,PRL,Italian}, we investigate the validity of the relations in Eq.  \ref{eq:eq1}. Laponite is a synthetic clay  that is widely used as a rheology-modifier in industrial materials such as paints, household cleaners and personal care  and house construction products. Furthermore,  it is considered as a model system for charged colloidal disks  and its  rich complex phase behavior  has attracted a lot of attention in recent years and has  been studied  extensively  from the fundamental point of view  as discussed in a recent review paper \cite{Lap-review}. Here, our aim is to find a link between slowing down of translational and rotational motion and increase of viscosity and investigate the possible development of dynamical heterogeneity in Laponite colloidal glass. The insights obtained from this study are possibly useful for other soft glassy fluids such as red blood cells or anisotropic colloids  \cite{RB}.

 At sufficiently high concentrations,  Laponite particles suspended in water, form a glassy phase that ages \cite{glass,PRL,Italian}. During aging the viscosity and  the relaxation times associated with translational and rotational relaxation increase \cite{Bonn1,Bonn2,Sara,MR-Sara}. The anisotropic shape allows us to study both the translational and rotational diffusion of the particles \cite{Pecora,Sara}.  Although  the aging of the translational and rotational diffusion has been already  investigated  in a previous work \cite{Sara},  their connection to the development of viscosity  in relation to  Stokes-Einstein and  Debye-Stokes-Einstein  relations is missing. The current study is  complemented with rheology and microrheology experiments. Thus we have  examined the problem from the more fundamental perspective of dynamical heterogeneity.  We  observe how the aging from a liquid-like to a disordered solid-like state influences the orientational and translational degrees of freedom, and how these are related to the increase of the global viscosity.
Further, by comparing the  translational diffusion
  of Laponite particles with that of a probe much bigger than their size, we can investigate how dynamical heterogeneity develops in the system as a function of waiting time.

\section{Materials and methods} The Laponite grade that we used for our experiments is Laponite XLG that consists of platelets of an average diameter of 25 nm and 1.2 nm thickness with an estimated polydispersity index of  about 30 \% \cite{Poly}. Laponite can absorb water, increasing its weight up to 20\%. Therefore, we first dried it in an oven at 100$^{o}$C for one week and subsequently stored it in a desiccator. Laponite dispersions are prepared in ultra pure Millipore water and are stirred vigorously by a magnetic stirrer 1.5 h to make sure that the Laponite particles carefully dispersed. The dispersions are filtered using Millipore Millex AA 0.8 $\mu$m filter units to obtain a reproducible initial state. This instant defines the zero of waiting time, $t_w = 0$ \cite{glass}.

Our  dynamic light scattering setup (ALV) is based on a He-Ne
laser ($\lambda = 632.8 $nm , 35 mW) and avalanche photodiodes as
detectors. An ALV-60X0 correlator directly computes the intensity
correlation functions
$g(q,t)=\frac{<I(q,t)I(q,0)>}{<I(q,0)>^{2}}$,  at  a scattering wave
vector number $q=\frac{4\pi n}{\lambda}\sin (\frac{\Theta}{2})$, in which
$\Theta$ is the scattering angle and $n$ is the refractive index of scattering sample.  For Laponite suspensions used in this study with a concentration around 3 wt\%, the refractive index is 1.335 \cite{refractive} and the measurements were always performed at angle $\Theta= \pi/2$, leading to a scattering wave number $q=1.87 \times 10^7$ m$^{-1}$.

The total electric field scattered by the colloidal particles with an axially symmetric optical anisotropy has a vertically polarized component $E_{VV}$ with an amplitude proportional to the average polarizability, $\gamma=(\gamma_{||}+2\gamma_{\bot})/3$ and a horizontal depolarized component $E_{VH}$ proportional to the intrinsic particle anisotropy $\beta=\gamma_{||}-\gamma_{\bot}$, the difference between the polarizabilities parallel and perpendicular to the optical axis \cite{Pecora}. Depolarized dynamic light scattering (DDLS) measures the correlation functions of the scattered light intensity whose polarization (horizontal) is perpendicular to the polarization of incident light (vertical), i.e., the VH mode, as opposed to the VV mode for which the polarization of scattered and incident light are both vertical.

The viscoelastic moduli during the aging process were also measured using a conventional Anton Paar Physica MCR300 rheometer with a Couette geometry. To avoid perturbing the sample during the aging process we performed the oscillatory shear measurements with a small strain amplitude of 0.01 at a fixed frequency 0.05 Hz. In order to prevent evaporation during the long time measurements, we installed a vapor trap. The experimental setup for performing microrheology consists of  optical tweezers formed by two  polarized laser beams. Details of this  experimental setup can be found in \cite{MRsetup}.

\section{Results and discussion}

We measured the VV and VH intensity correlation functions  at a fixed scattering angle $\Theta=90^{o} $ regularly  during the aging of glassy suspension.  The  VV intensity correlation functions  reflect mainly the dynamics of translational degree of freedom as the contribution of the rotational motion to the VV correlations  is rather small as has been  demonstrated before \cite{Sara}. The VH correlation functions, on the other hand are determined by both translational and rotational degrees of freedom. Indeed, for dilute enough suspensions, the VH correlations  can be written as  a product of  VV correlation function and a purely orientational correlation  \cite{Pecora,Sara}.

\begin{figure}[h!]
\includegraphics [scale=0.6]{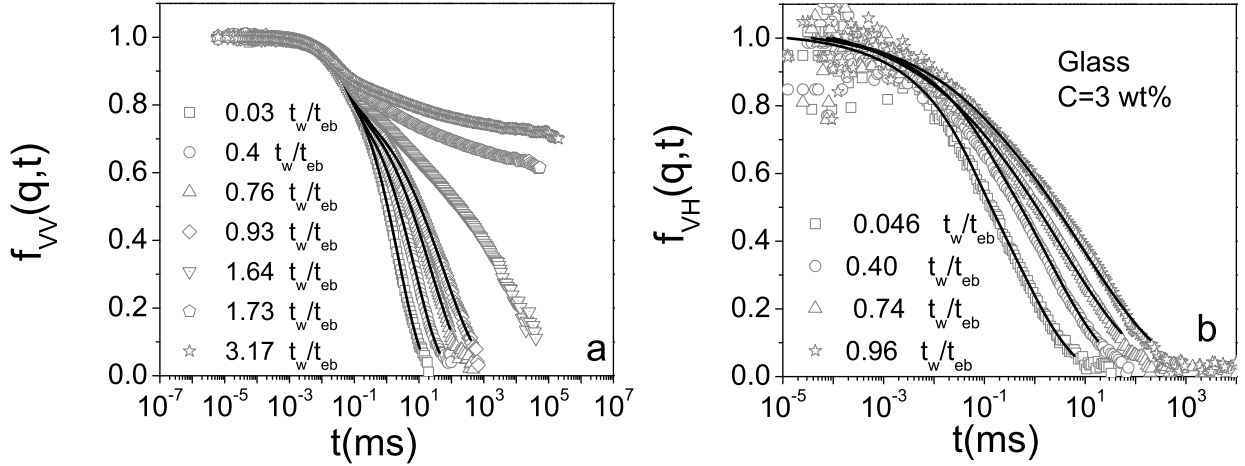}
\caption{
 Evolution of polarized (VV) and depolarized (VH) intermediate scattering functions (symbols) and their corresponding fits (solid lines) for a glass (Laponite 3 wt\%, pure water). The waiting times are shown in the legends in terms of the ergodicity-breaking time $t_{eb}=450$ min. The last three VV data-sets are in the non-ergodic state obtained by  ensemble-averaged measurements.}\label{fig1}
\end{figure}

Studying the aging dynamics of VV correlations for an extensive range of samples, it has been well established that one can observe two regimes of aging \cite{kroon,glass,PRL}. In the first regime of aging, the translational correlation functions decay to zero within the experimental accessible time scales and the average relaxation time grows exponentially with waiting time. In  the second regime of aging,  the ensemble-averaged correlations do not decay to zero and a plateau in intermediate scattering function appears (see Fig. \ref{fig1}a). The waiting time for which the correlation functions
(i) no longer decay to zero within the experimental time-scale and
(ii) the time-averaged correlation functions are not equal to their ensemble-averaged values
defines the ergodicity-breaking time $t_{eb}$. Studying the aging dynamics of VV and VH correlation functions simultaneously, we find out that the VH correlations become non-ergodic at nearly the same time as VV correlations. Therefore,  we take the same $t_{eb}$ value for both VV and VH correlations. To be able to quantify the relaxation times for rotational and translational diffusion, here we only focus on aging dynamics in the first, ergodic, regime of aging.

In Fig. \ref{fig1}, we show the $f_{VV}$ and $f_{VH}$ intermediate scattering functions extracted from intensity correlation functions measured at different stages of aging. We already know that for the VV intermediate scattering functions $f_{VV}$ a two-step relaxation can be observed \cite{Bonn1}; we observe a similar
relaxation for VH correlations as well \cite{Sara}.  In the ergodic
regime, we can quantify the relaxation times by fitting the VV
intermediate scattering functions  by a sum of an exponential and a
stretched exponential  \cite{Bonn1}.
\begin{equation}\label{eq:eq2}
f(q,t)=A \exp(-t/\tau_{1})+(1-A)\exp(-(t/\tau_{2})^{b}).
\end{equation}
where $A$ determines the relative contribution of exponentially decaying fast relaxation time $\tau_1$ and $b$ is known  as the stretching exponent characterizing the broadness of  slow relaxation modes  that contribute to the correlation function. The smaller $b$ is, the broader is the distribution of the slow relaxation times. $\tau_2$ gives us an idea about the mean relaxation time that is determined as $\tau_m= \tau_2/b \Gamma (1/b)$, where $\Gamma$ is the gamma function.

 The VV  mode data reflects the aging of the translational degree of freedom, and the
relaxation times from the fit are therefore a direct measure of the
relaxation times of translational diffusion. Fitting the VV correlations with Eq. \ref{eq:eq2}, we find that $A$  is constant with waiting time with a value $0.215 \pm 0.02$. The same holds for the fast relaxation time whose inverse gives us the short-time diffusion  coefficient $D_s$, i.e. $\tau_1=1/ (D_s q^2)$. As was shown in previous works \cite{Abou,Sara}   $\tau_2$ is growing  exponentially with waiting time concomitant with the decrease of the stretching exponent $b$  whose value decreases linearly from  $0.606 \pm 0.005$ at $ t_w \approx 0$ to $0.358 \pm 0.005$ at $ t_w \approx t_{eb}$. In  Fig. \ref{fig3}a,  we have presented  the evolution of both fast ($\tau_1 $) and slow ($\tau_2$)  relaxation times; the fast relaxation time is usually interpreted as a 'cage rattling' motion, whereas the longer relaxation time corresponds to a cage reorganization.

As mentioned before, both translational and rotational degrees of freedom contribute to the
VH correlations. In order to gain a more direct insight into the
rotational dynamics, we extract the orientational correlation
functions defined as the ratio $f_{or}=f_{VH}/f_{VV}$ as shown in Fig. \ref{fig:orientation}. These orientational correlations also can also be fitted with the sum of a single and a stretched exponential according to Eq \ref{eq:eq2}. The corresponding  fast $\tau_1 \propto 1/D^{s}$ and slow $\tau_2$  rotational relaxation times can be extracted as depicted in Fig. \ref{fig3}b. {We find that for the orientational correlations,  $A$ decreases with waiting time from $A=0.12 \pm 0.03$ at early stages of aging to $A=0.012 \pm 0.01$ at $t_w \approx t_{eb}$. The stretching exponent $b$
as well decreases from $b=0.30 \pm 0.01$ at early stages of aging to $b=0.15 \pm 0.02$ upon approaching the ergodicity-breaking time, pointing to a broader distribution of rotational slow relaxation times as compared to the translational ones.

\begin{figure}[t]
\begin{center}
\includegraphics [scale=0.4]{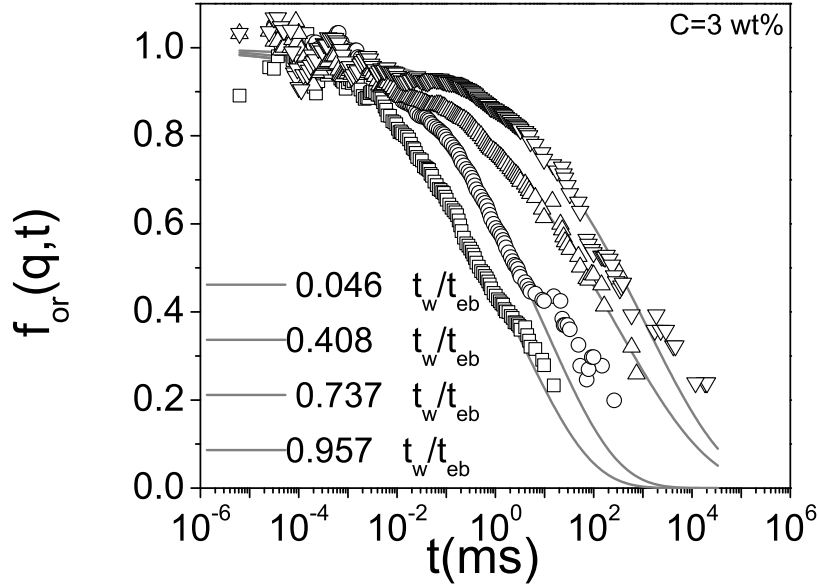}
\caption{
 The orientational correlation functions defined as $f_{VH}/f_{VV}$
 at different waiting times in
a glass (Laponite 3 wt\%, pure water).  The lines show the fits with the sum of a single
and a stretched exponential according to Eq. \ref{eq:eq2}.
  }\label{fig:orientation}
 \end{center}
\end{figure}

We also measured the macroscopic  complex shear modulus with the rheometer; the complex viscosity magnitude can be obtained as $|\eta^*|=\sqrt{G'^2+G''^2}/\omega$. With these measurements, we are then in a position to discuss the translational and rotational relaxation times and compare them to the measured macroscopic viscosity.

As can be observed from Fig. \ref{fig3} there are important differences between the aging rates of the different  measured quantities, i.e., translational and rotational diffusion and viscosity}.  The first important observation is that the slow relaxation time of the orientational degree of freedom grows faster than the corresponding one for the translational degree; this is the first indication of the decoupling of translational degree of freedom from the rotational degree during the aging of Laponite suspensions. The second important observation comes from the comparison of the light scattering results with the low-frequency complex viscosity modulus as a function of waiting time, as also shown in Fig. \ref{fig3}.

\begin{figure}[t]
\begin{center}
\includegraphics [scale=0.3]{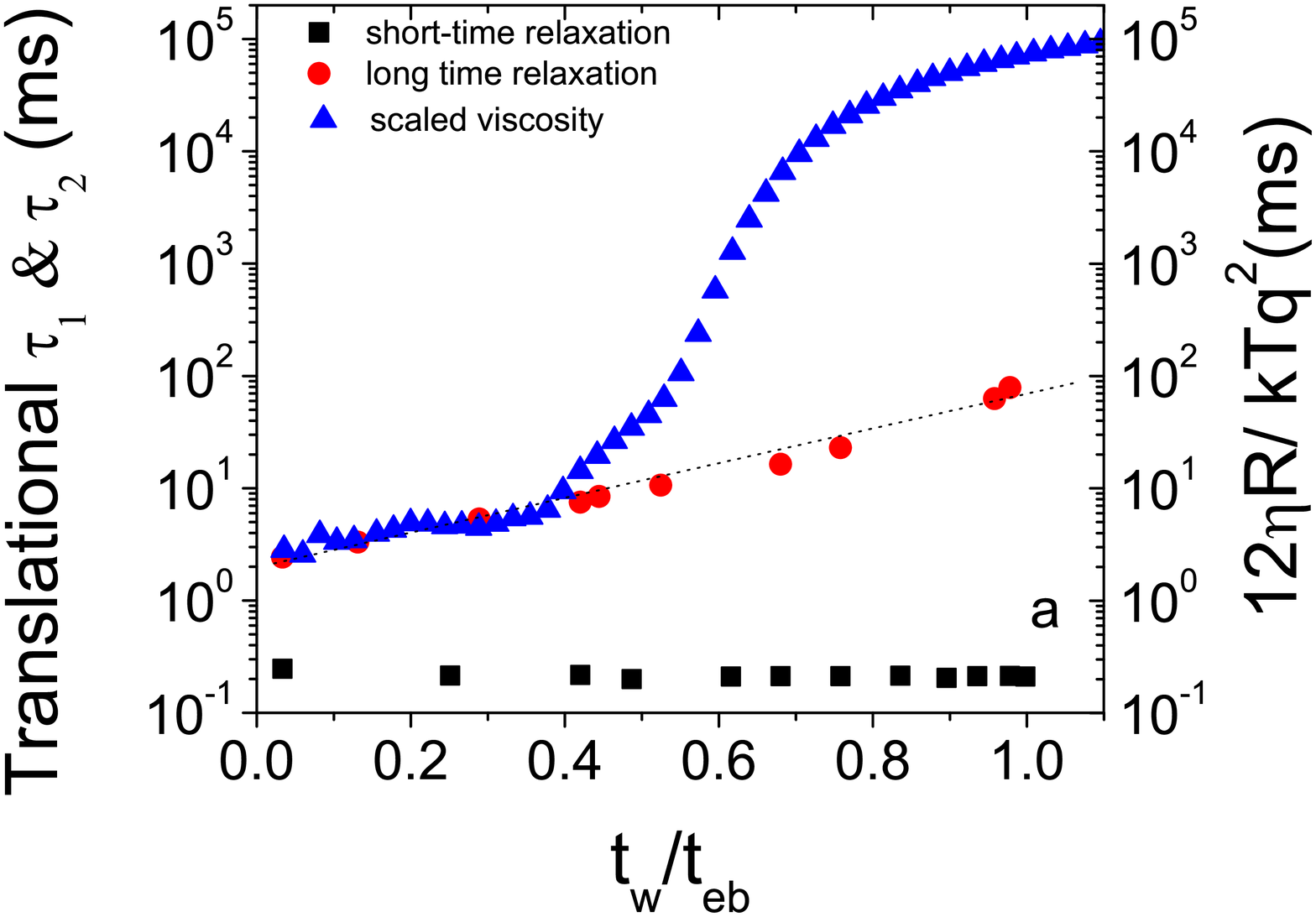}
\includegraphics [scale=0.3]{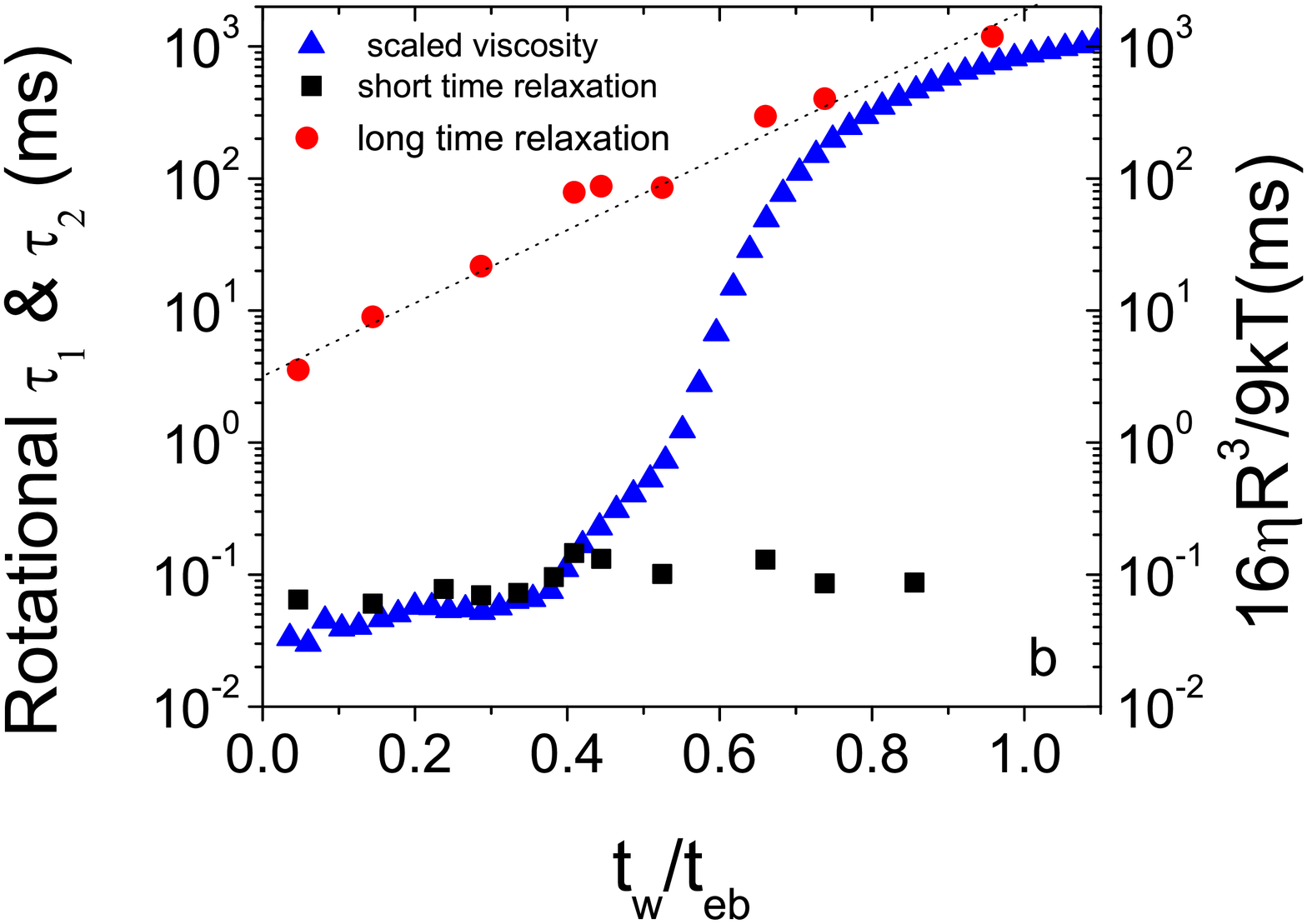}
\caption{
 The   fast and slow translational and orientational relaxation times and  the scaled viscous modulus at $f=0.05$ Hz versus scaled waiting time
 measured for a Laponite suspension of $C= 3$ wt\% (glass). The viscosity is scaled in such a way that if Stokes-Einstein relation $D_{tr}=\frac{k_BT}{ 12 \eta R}$  and Stokes-Einstein-Debye relation $D_{r}=\frac{3 k_BT}{ 32 \eta R^3}$ were valid, one would obtain the translational (rotational) relaxation time from the corresponding viscosity. The dashed lines shown are just for a guide to the eyes.
  }\label{fig3}
 \end{center}
\end{figure}

 In Fig. \ref{fig3}a (b) the viscosity is scaled in such a way that if Stokes-Einstein(-Debye) relation were valid, one would obtain the translational (rotational) relaxation time from the corresponding viscosity according to the Stokes-Einstein (-Debye) relation. From this figure, we interestingly find that the growth  of both translational and rotational relaxations occurs at a different rate than that of viscosity. In other words, not only translational and rotational degrees are decoupled but also the structural relaxation time of the system characterized by viscosity is decoupled from these degrees of freedom.  Looking at Fig.  \ref{fig3}, we can recognize three regimes of aging for the evolution of viscosity. The first regime of aging ($t_w< 0.4t_{eb}$ ) where the viscosity  grows at a similar rate as that of the slow translational relaxation time and the Stokes-Einstein relation is obeyed but the slow rotational relaxation time grows at a much faster rate, although short-time rotational diffusion shows a good agreement with the viscosity.  The second regime ( $0.4t_{eb}<t_w< 0.75 t_{eb}$) where the viscosity growth goes beyond the slow translational relaxation time and the slow rotational relaxation time still remains much higher than the other two relaxations. In the third regime ($t_w> 0.75 t_{eb}$), Upon approaching the ergodicity-breaking time, the structural relaxation time becomes comparable to the slow rotational relaxation time, however the slow  translational relaxation time still remains considerably lower, signaling an enhanced translational diffusion compared to rotational diffusion and structural relaxation.

The observed behavior in Fig. \ref{fig3} represents the central result of our work and  can be understood in light of spatially heterogenous dynamics, i.e., formation of  mobile regions in which the  translational (rotational) dynamics of particles are correlated \cite{decoupDynHet}. The observation of a stretched exponential for both translational and orientational correlations  points to a broad distribution of relaxation times supporting the existence of underlying spatially heterogenous domains. The decrease of the stretching exponents with waiting time \cite{Sara} shows that the width of distribution of relaxation times  also increases with waiting time. Particularly, the different behavior of translational and the rotational correlations is very interesting and shows that the rotational degree of freedom becomes glassy at a  faster rate. This points to the fact that  translationally correlated mobile regions do not necessarily correspond to the domains where the rotational motion of particles are correlated. Indeed, recent measurements of a monolayer of ellipsoidal particles have revealed that translational and rotational cooperative motions are anti-correlated in space and  translationally mobile regions correspond to  pseudo-nematic domains where the orientational degree of freedom is frozen \cite{OrientationalG}.

The much slower growth of the translational relaxation time compared to rotational and structural ones points to an enhanced translational diffusion that becomes more pronounced with increasing waiting time. However, one should note that the average  translational diffusion is not simply the inverse of the average relaxation time because of a broad distribution of relaxation times. The average translational diffusion that involves the passage of particle through different domains and is  determined by $\langle \tau^{-1} \rangle$  rather than  $1/\langle \tau \rangle$  \cite{Tarjus}.

 A  more quantitative picture for the translational diffusion can be obtained by extracting the frequency-dependent diffusion  from dynamic light scattering measurements. The frequency-dependent diffusion can be obtained from Laplace transform of intermediate scattering function according to the following relation \cite{Klein}:
  \begin{equation} \label{DW}
 S(q,\omega)=\int_0^{\infty} f(q,t) \exp(i\omega t)=\frac{1}{-i\omega+D(q,\omega)q^2}
 \end{equation}

  Fitting $f(q,t)$ with Eq. \ref{eq:eq2}, we can obtain its Laplace transform analytically using Mathematica and therefore extract $D(q,\omega)$.

We can now compare this frequency-dependent diffusion coefficient with that of a probe particle whose size is much larger than Laponite particles. Such large particles probe the global viscosity, and it has been shown previously that for a probe size much larger than the Laponite size, the diffusion coefficient of the probe obeys the Stokes-Einstein relation \cite{PRL-Sara}: it  gives  frequency-dependent complex viscosity that is equal to the bulk viscosity \cite{MR-Sara}.  Utilizing the microrheology technique, we obtained the displacement fluctuations of  a  micron-sized probe particle trapped by optical tweezers and  directly computed the power spectrum of displacements, $D(\omega)=\frac{1}{2}\omega^2 |x(\omega)|^2$.

 In order to compare the diffusion coefficients from Microrehology and DLS experiments, we have
scaled their values with the factors $6\pi R_{probe}$ and $12R_{Laponite}$,
respectively to remove the trivial dependence of diffusion
coefficient to particle size and shape. In Fig. \ref{fig5}, we have
depicted the frequency-dependent diffusion of Laponite particles
and the probe particle at two different waiting times. At early
stages of aging the two diffusion coefficients agree at high
frequencies, while the diffusion of Laponite particles is
significantly lower at low frequencies.  Interestingly, this is in line with our earlier observation of the agreement of the short-time rotational diffusion with the viscosity at early stages of aging. For small $t_w$, the viscosity is frequency-independent \cite{MR-Sara}; therefore the measured viscosity at high frequencies has the same value as that measured at $f=0.05$ Hz.  This suggests that for high frequencies both  Stokes-Einstein and Stokes-Einstein-Debye relations hold at the early stages of aging. As the sample ages, this difference becomes stronger and the two diffusion coefficients no longer coincide at any frequency.  This demonstrates that the diffusion of the larger particle, i.e. the viscosity, evolves at a faster rate than the Laponite particles themselves, in line with the results presented in Fig. \ref{fig3}.

 \begin{figure}[h!]
\begin{center}
\includegraphics[scale=.6]{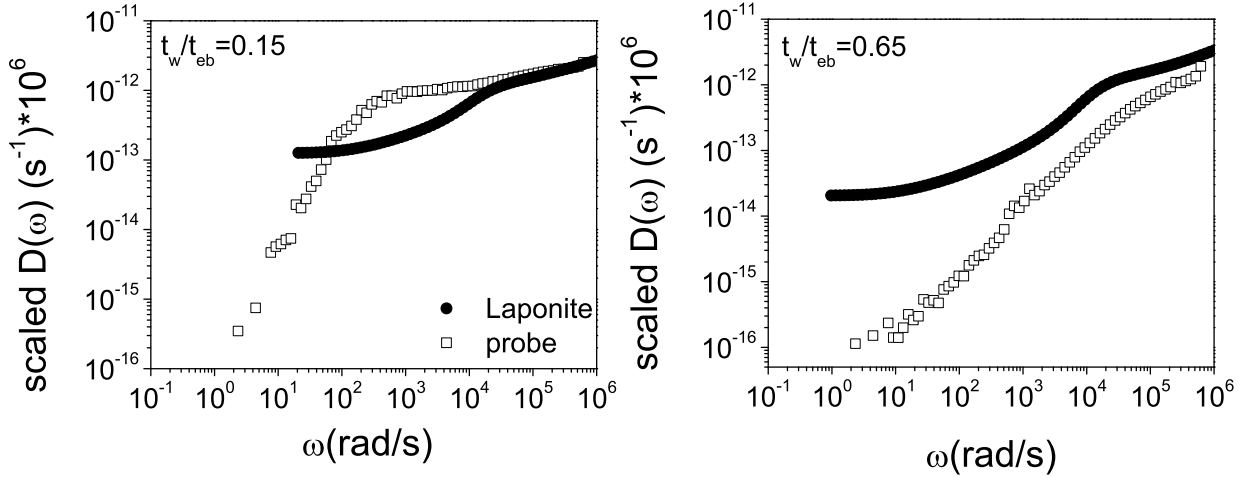}
\caption{The frequency-dependent diffusion coefficient of Laponite
particles at a scattering vector corresponding to $qR=0.3$
compared to the diffusion of a probe particle of diameter 1.16
$\mu m$ at two different stages of aging in a colloidal glass of
Laponite 3.2 wt \%. The waiting times are shown in the legend.
Note the diffusion of probe and Laponite particles are scaled with
their radii to be comparable.} \label{fig5}
\end{center}
\end{figure}
 To demonstrate  the  scale-dependence of  diffusion  with waiting time, in Fig. \ref{fig5}, we have plotted the the ratio of  scaled diffusion coefficient of Laponite particles  to scaled diffusion coefficient of probe particle as a function of waiting at a low $\omega=21$ (rad/s) and a high frequency   $\omega=6 \time 10^6$ (rad/s). As can be seem from this figure this ratio is close to 1 at high frequencies while it grows with waiting time for the low frequency. The comparison between the two methods clearly  shows the development  of a scale-dependent diffusion coefficient, which in turn demonstrates the development of a spatially heterogenous dynamics that becomes more pronounced as the waiting (aging) time increases. The microrheology measurements, performed with probe particles of diameters of 0.5 and 1 micrometer, showed that beyond 0.5 micrometer the translational diffusion of probe particles  no longer shows a  size-dependence and the Stokes-Einstein relation is obeyed: the local viscosity measured by micron-sized particles agrees with that macroscopic viscosity \cite{MR-Sara}. This means that at this length scale the probe is large and slow enough to average out the influence of dynamical heterogeneity and consequently the Stokes-Einstein relation is valid \cite{PRL-Sara, MR-Sara}. Our findings also agree with other experiments that investigated the diffusion of tracer particles of different sizes as a function of waiting time \cite{size-dependentD,size-dependentD1}. Measuring the diffusion of 50, 100 and 200 nm size tracer particles in aging Laponite suspensions, Strachan et. al. found that their scaled diffusion coefficients are
identical at the beginning of aging, however as the time proceeds, the correlation functions of the larger tracer particles evolve at a faster rate \cite{size-dependentD}. Also in line with our findings,  Petit et. al. \cite{size-dependentD1} investigated the dependence of translational diffusion on probe size at late stages of aging and report that the diffusion coefficient becomes independent of probe size for length scales larger than the interparticle distance.

 \begin{figure}[h!]
\begin{center}
\includegraphics[scale=.3]{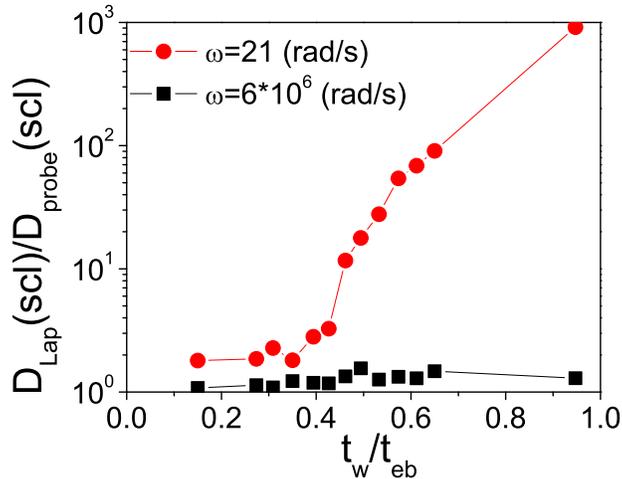}
\caption{ The ratio of the frequency-dependent diffusion coefficient of Laponite
particles  to the diffusion of a probe particle of diameter 1.16
$\mu m$ at two different frequencies  (shown in the legend) as a function of waiting time during the aging in a colloidal glass of
Laponite 3.2 wt \%. Note the diffusion of probe and Laponite particles are scaled with
their radii to be comparable.} \label{fig5}
\end{center}
\end{figure}

The developments of scale-dependent diffusion at later stages of aging shows that the glass formation is accompanied by a growing dynamic correlation length scales during the aging  similar to the trend found in hard sphere glass suspensions upon increasing the volume fraction \cite{growingL}.
  The  length scale beyond which there is no scale-dependent diffusion can provide us an estimate of the size of the  spatially correlated mobile regions in the sample at late stages of aging.  Our  results  set an upper bound  of 500 nm for the dynamic correlation length. On the other hand,
  results  of \cite{size-dependentD,size-dependentD1}  give us a rough estimate for the size of dynamical correlation length $100  \leq \xi_d < 200$ nm which is between 3 to 7 particle diameters. Of course, a more systematic study of probe size-dependent diffusion is required to determine the exact waiting time-dependence of such a dynamical correlation length.

The growth of the average rotational relaxation time concomitant with the broadening of the width distribution of relaxation times  suggests that there should be a distinct dynamic correlation length associated with the characteristic slow time for the  orientational degree of freedom, that also grows with increasing waiting time. Interestingly, our data show that the orientational relaxation time at late stages of aging agrees with the macroscopic viscosity.  How the orientational relaxation depends on the size of probe particle and how the translationally and rotationally mobile domains develop with waiting time, is an issue that merits further investigations. However, our data do support the existence of dynamically heterogenous domains for both degrees of freedom that are not strongly correlated.

At this point it is instructive to compare our results of anisotropic  colloidal glass with findings on molecular glass formers \cite{decoupling1,decoupling2,decoupDynHet} where both translational and rotational relaxations grow  concomitant with the increase of viscosity upon supercooling. Studies on supercooled  liquids have found out that the rotational correlation time follows the temperature dependence of  the Debye-Stokes-Einstein relation to within a factor of 2 or 3 while the viscosity changes by 12 orders of magnitude \cite{decoupling1,decoupling2} upon cooling. To the contrary, the translational diffusion does not  follow the temperature-dependence expected from Stokes-Einstein relation \cite{decoupling1,decoupling2,decoupDynHet} upon decreasing the temperature. Similarly to what happens for our colloidal glass, in these systems  (i) a faster translational diffusion compared to rotation and
(ii) a dependence of the diffusion coefficient on probe-size
have been observed  \cite{decoupDynHet}.
Therefore, our observations are qualitatively in agreement with findings on molecular glass-formers, although as we study here the waiting-time dependence rather than temperature dependence, some features are particular to our aging colloidal glass.

To summarize, the observed decoupling between rotational and translational relaxation times and viscous relaxation, can be understood in terms of
 dynamical heterogeneity. Particularly, our data suggest that the development of heterogenous domains for translational and  orientational degree of freedom are distinct in accord with the recent finding on a glass of ellipsoidal colloidal suspensions \cite{OrientationalG}.
 The observed  behavior of rotational diffusion at early stages of aging and its strong decoupling from translational and structural relaxation is  different from the reported measurements in the literature and requires further experimental and theoretical investigations.

\textbf{Acknowledgments}
We would like to thank Giulio Biroli and Gilles Tarjus for stimulating and fruitful discussions. This research has been supported
by the Foundation for Fundamental Research on Matter (FOM), which is financially supported by Netherlands Organization for
Scientific Research (NWO). S. J-F. was further supported by the foundation of "Triangle de la Physique".  LPS de l'ENS is UMR8550 of the CNRS, associated with the universities Paris 6 and 7.

\end {document}